\documentclass[twoside]{ilcws10}
\usepackage[latin1]{inputenc}
\usepackage[dvips]{graphicx,epsfig,color}
\usepackage{wrapfig,rotating}
\usepackage{amssymb,amsmath,array}

\begin{document}
\title{
%%%%   Paper title goes here  %%%%%%%%%%%%%%
%Measurement of the Higgs mass via the 
%channel\\
MEASUREMENT OF THE HIGGS MASS VIA THE CHANNEL: 
%$e^{+} e^{-} \to Z H \to e^{+}e^{-} + X$} 
e$^{+}$e$^{-} \to~$Z H $\to~$e$^{+}$e$^{-}$ + X}
% (PRELIMINARY RESULTS)}%%
%***********************************************************************
% AUTHORS INFORMATION AREA
%***********************************************************************
\author{D. Benchekroun$^1$, J-Y. Hostachy$^2$, Y. Khoulaki$^1$ and L. Morin$^2$
% Optional short acknowledgment: remove next line if non-needed
\thanks{Research conducted in the scope of a LIA known as 
"International Laboratory for Collider 
Physics - ILCP". 
The moroccan contribution to this work is also 
supported by the High Energy Physics Network (RUPHE).
}
% DO NOT MODIFY THE FOLLOWING '\vspace' ARGUMENT
\vspace{.3cm}\\
% Addresses and institutions (remove "1- " in case of a single institution)
1- Université Hassan II, Faculté des Sciences Aïn Chock,\\
BP 5366 Maârif, Casablanca - Morocco
%% Remove the next three lines in case of a single institution
\vspace{.1cm}\\
2- Laboratoire de Physique des Subatomique et de Cosmologie (LPSC),\\ 
Université Joseph Fourier (Grenoble I), CNRS/IN2P3,\\ 
Institut Polytechnique de Grenoble,\\
53 rue des Martyrs, F-38026 Grenoble Cedex - France\\
}
%%***********************************************************************
% END OF AUTHORS INFORMATION AREA
%***********************************************************************

\maketitle

\begin{abstract}
In this communication, the mass declined for the decay channel, 
$e^{+} e^{-} \to Z H \to e^{+}e^{-} + X$, as
measured by the ILD detector was studied. 
The Higgs mass is assumed to be 120~GeV 
and the center of mass energy is 250 GeV.
For an integrated luminosity of 250~fb$^{-1}$,
the accuracy of the reconstruction
and the good knowledge of the initial state allow for the measurement of the
Higgs boson mass with a precision of about 100~MeV.\\
\end{abstract}

\section{Introduction}

%\end{figure}
One of the major goals of the ILC ("International Large Collider") is 
to observe the Higgs boson and describe its properties. 
In the Standard Model~\cite{bibli:modeleWS} the Higgs particle plays a key role 
\begin{wrapfigure}{r}{0.5\columnwidth}
%\begin{figure}
\centerline{\includegraphics[width=0.5\columnwidth,height=4cm]{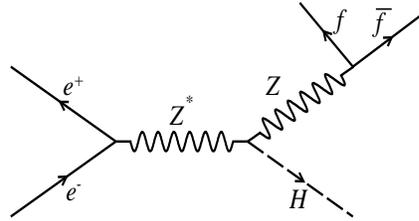}}
\caption{Higgs boson production via Higgs-strahlung process.}
\label{Fig:Higgs-strahlung}
\end{wrapfigure}
in explaining the origin of the masses of the elementary particles. 
If existing, a Higgs boson with a mass of 120 GeV is favoured by recent 
analyses of electo-weak data~\cite{bibli:resutats-LEP}.
The Higgs-strahlung process ($e^{+} e^{-} \to Z^{*} \to Z H$)  
is the major Higgs production mechanism at ILC, 
see Figure~\ref{Fig:Higgs-strahlung}.
In particular, in this study, we will consider the channel 
$e^{+} e^{-} \to Z H \to e^{+}e^{-} + X$.

By detecting the $e^{+} e^{-}$ lepton 
pair produced by the well-known Z boson decay, one can 
measure the Higgs mass (using the mass recoiling of the Z) 
independently of the Higgs decay. 
The most favourable case  
is obtained for the center of mass energy
slightly larger than the Z boson mass plus the Higgs boson mass 
(i.e. $\gtrapprox$ 211.2~GeV) \cite{bibli:Li2}. 

\section{Event generation and detector simulation}

The analysis was performed with the International Large detector (ILD) 
data sample 
fulfilled for ILD Letter of Intent~\cite{bibli:LoI}.
The events are fully simulated and 
reconstructed with the 
ILD\_00 detector model.
The Higgs mass is assumed to be 120 GeV, and the center 
of mass energy is 250 GeV, with an integrated luminosity of 250~fb$^{-1}$.
The e$^{+}$ polarization is taken to be equal to 30\% and the   
e$^{-}$ polarization is 80\%.\\
\begin{wraptable}{l}{0.5\columnwidth}
\centerline{\begin{tabular}{|c||c|c|}
   %\cline{2-5}
   \hline
   \multicolumn{1}{|c||}{Cross section } & \multicolumn{2}{c|}{e$^{+}$e$^{-}$ beam polarization}\\
   \multicolumn{1}{|c||}{ (fb) } & \multicolumn{2}{c|}{mode (30\%, 80\%)}\\
   \hline
   Process & $(+,-)$ & $(-,+)$ \\
   \hline
   ZH$\to$eeX & 12.55 & 8.43 \\
   \hline
   ee (Bhabha) & 17.30 10$^{6}$ & 17.28 10$^{6}$ \\
   \hline
   4f$\to$eeff & 4897 & 3793 \\
   \hline
\end{tabular}}
\caption{ZH$\to$eeX and background cross sections  
as a function of the beam polarization.}
\label{tab:pola}
\end{wraptable}
Table~\ref{tab:pola} gives the signal (ZH$\to$eeX)
cross section as a function of the 
beam polarization (e.g. (+,-) mode means that the positron beam
polarization is +30\% and the electron beam polarization is -80\%).
The polarization modes (-,-) and (+,+) lead to smaller signal cross 
sections: 7.65 and 5.78 fb respectively for similar levels of background.
The Bhabha scattering (ee) and the Standard Model 
events with 4 fermions, including e$^{+}$e$^{-}$ contribute to 
the background. 
The cross section for Bhabha scattering is
6 orders of magnitude larger than the signal cross section.

\section{Event reconstruction and background rejection}
The identification of the Z boson is obtained by selecting 
the e$^{+}$e$^{-}$ pair which gives the best mass for the Z boson.
In our case: 
M$_{Z reconstructed}$ = M$_{Z} \pm$ 10~GeV. 
In addition, central leptons with opposite charge are required, i.e.: 
$\vert$cos($\theta_{i}$)$\vert$ $<$ 0.9 
(in that case the momentum measurement is optimal)

\begin{wraptable}{l}{0.5\columnwidth}
\renewcommand{\arraystretch}{1.2}
\centerline{\begin{tabular}{|c|c|c|c|c|}
\hline
{\tiny (+,-)} & {\tiny L$_{simul}$} & {\tiny N$_{simul}$} & {\tiny N$_{reconst}$} & {\tiny N$_{expected}$}  \\
 & {\tiny (fb$^{-1}$)} & & & {\tiny for 250 fb$^{-1}$} \\
\hline
{\tiny eeX} & {\tiny 10 000} & {\tiny 125 500} & {\tiny 49 799} & {\tiny 1 245} \\
\hline
{\tiny ee} & {\tiny 0.5123} & {\tiny 8 866 734} & {\tiny 48 201} & {\tiny 23 10$^{6}$} \\
\hline
\end{tabular}}
%\caption{Expected number of events.}
\caption{Number of simulated, reconstructed and expected events 
with their associated luminosity for the $(+,-)$ beam polarization mode.}
\label{tab:result1}
\end{wraptable}

The Higgs mass is then calculated from the well known formula:\\
\hglue 1.5cm
 M$_{H}^{2}$ = s + M$_{Z}^{2}$ - 2M$_{Z}\sqrt{s}$,\\
where M$_{Z}$ is the reconstructed mass of the Z boson and $\sqrt{s}$ 
the total energy in the mass center of the collision.\\
The expected number of events after selection is given in 
Table~\ref{tab:result1}.
Despite a worthy gain,
the Bhabha effect remains strongly dominant.\\
\section{The Bhabha scattering}
\begin{wrapfigure}{r}{0.5\columnwidth}
\vglue -10.5cm
\centerline{\includegraphics[width=7cm,height=6cm]{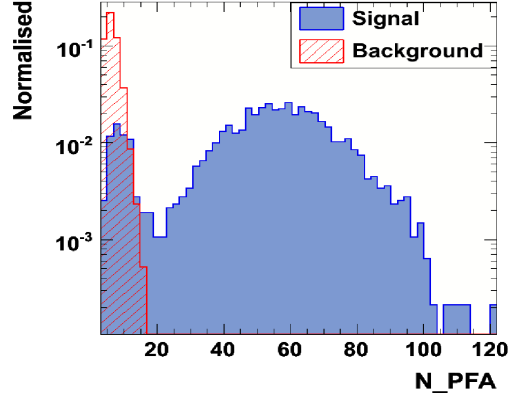}}
\caption{Normalized distributions (surfaces equal to 1) of the reconstructed 
objects from the Particle Flow 
Algorithm (PFA) \cite{bibli:PFA} for the ZH signal (in blue) and the ee Bhabha 
scattering (in red).}
\label{Fig:N_PFA}
\end{wrapfigure}

To reject the Bhabha events, a cut on 
the number of reconstructed objects (N$_{objects} >$ 21) is applied, 
see Figure~\ref{Fig:N_PFA}. 
This cut removes the first maximum in the signal distribution 
which probably corresponds to events where the Higgs boson 
decays into a ${\tau}^+$${\tau}^-$ pair. 
Unfortunately, the measurement of the cut efficiency is 
limited by the available Monte Carlo event number, see Table~\ref{tab:result2}.
Nevertheless, this cut 
looks very efficient because of the shape of the Bhabha distribution.
Therefore, in the following parts the Bhabha effect will be neglected but 
the cut on the number of reconstructed objects 
will be maintained.

\vglue 2.5cm
\begin{wraptable}{l}{0.45\columnwidth}
\centerline{\begin{tabular}{|c|c|c|c|}
\hline
 & & & \\
{\small (+,-)} & {\small L$_{simul}$} & {\small N$_{objects} >$ 21} & {\small N$_{objects} >$ 21} \\
 & {\small (fb$^{-1}$)} & & {\small for 250 fb$^{-1}$} \\
 & & & \\
\hline
 & & & \\ 
{\small eeX} & {\small 10 000} & {\small 55 847} & {\small 1396} \\
 & & & \\
\hline
 & & & \\
{\small ee} & {\small 0.5123} & {\small 1} & {\small 487} \\
 & & & \\
\hline
\end{tabular}}
\caption{Number of events, with their associated luminosity,after the cut on the number of reconstructed objects 
for the $(+,-)$ beam polarization mode.}
%for the $(+,-)$ beam polarization mode.}
\label{tab:result2}
\end{wraptable}

\section{Event selection (cuts on kinematic variables)}
\begin{figure}[tbp]
\scalebox{1}{

\begin{tabular}{cc}
\includegraphics[width=6.7cm,height=4cm]{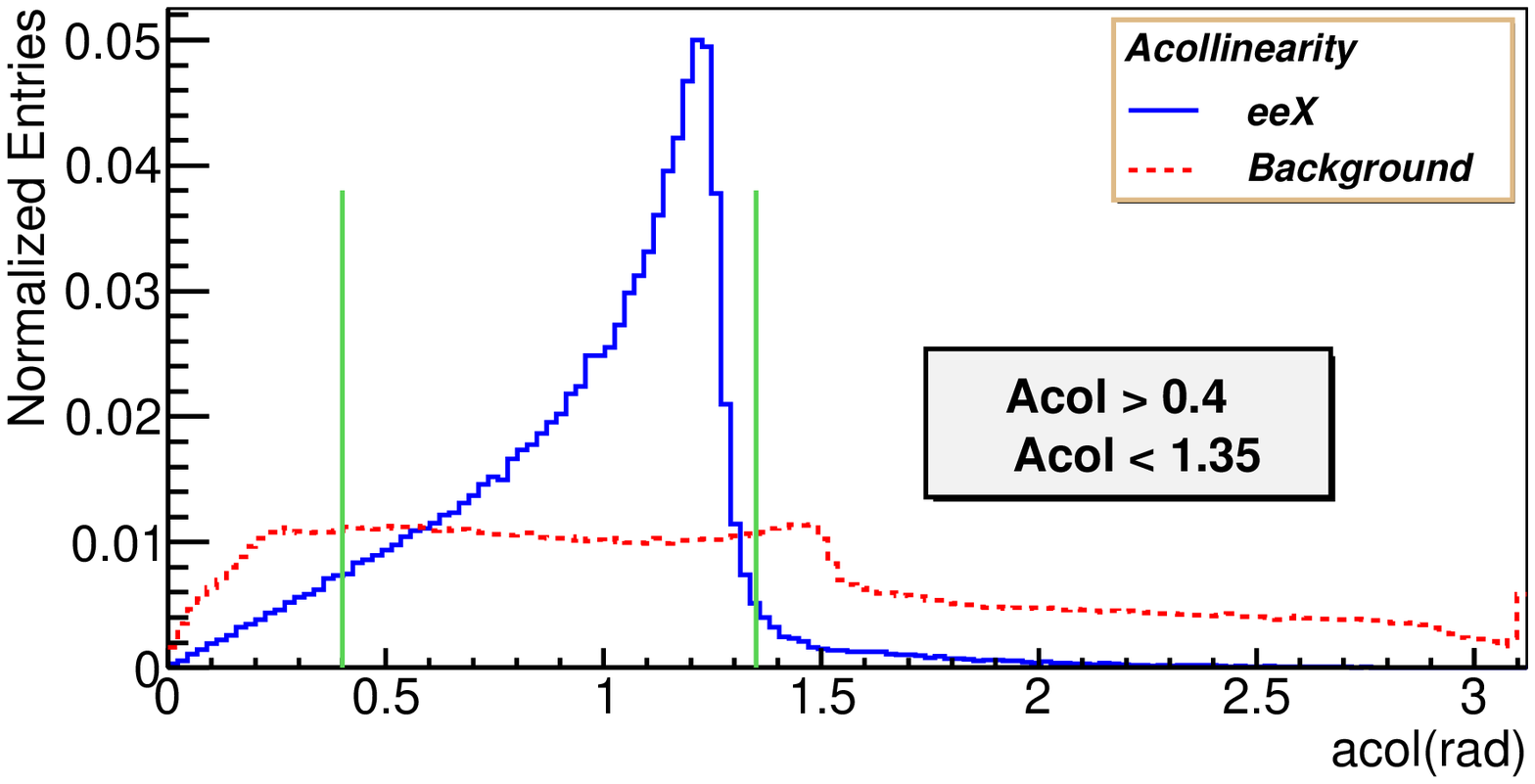} &
\includegraphics[width=6.7cm,height=4cm]{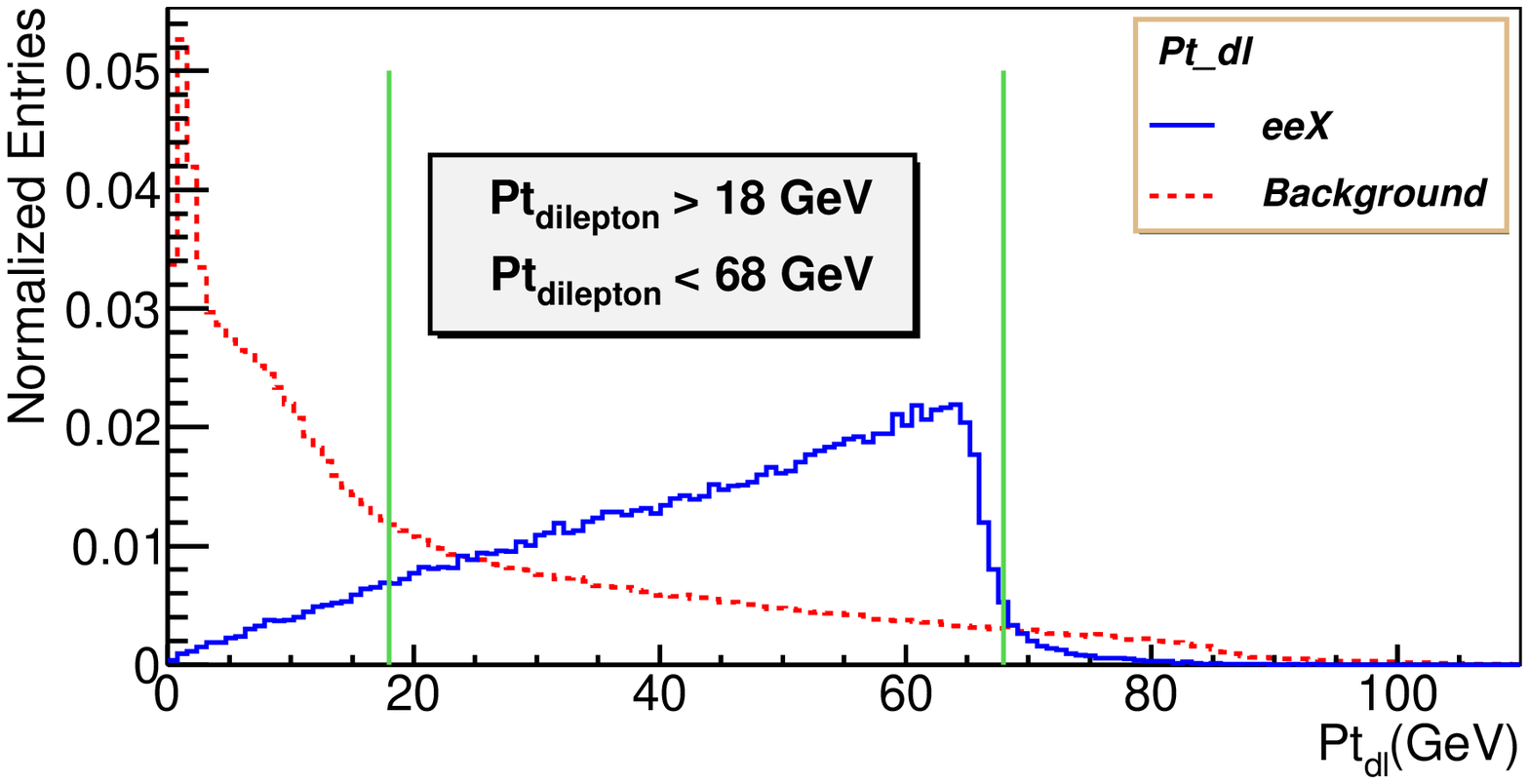}\\
\includegraphics[width=6.7cm,height=4cm]{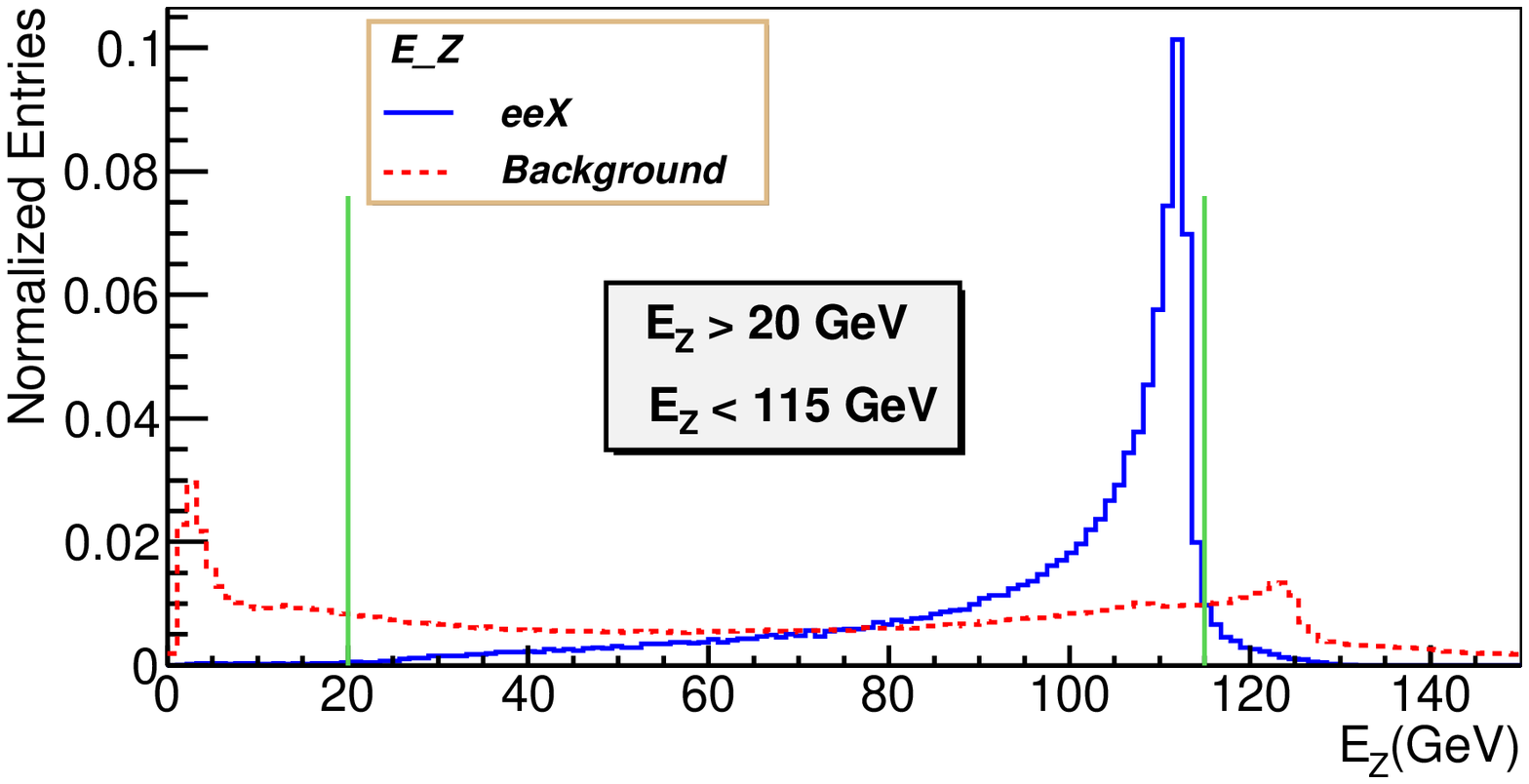}&
\includegraphics[width=6.7cm,height=4cm]{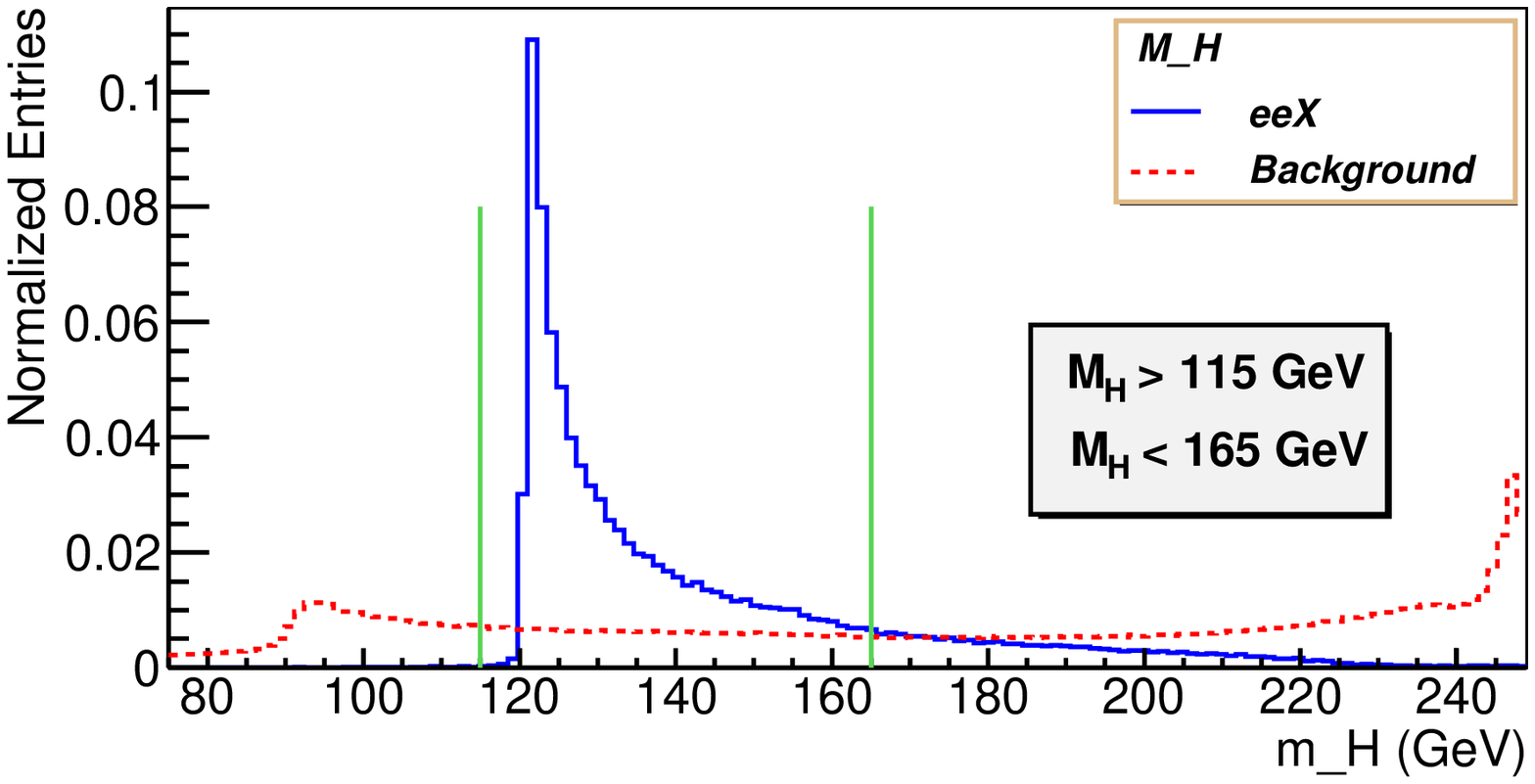}\\
\end{tabular}}
\caption{Event selection cuts (the distribution surfaces are normalized to 1).}
\label{Fig:cuts}
\end{figure}

\begin{figure}[tbp]
\resizebox{\textwidth}{!}{
\begin{tabular}{cc}
\includegraphics[width=6.7cm,height=4cm]{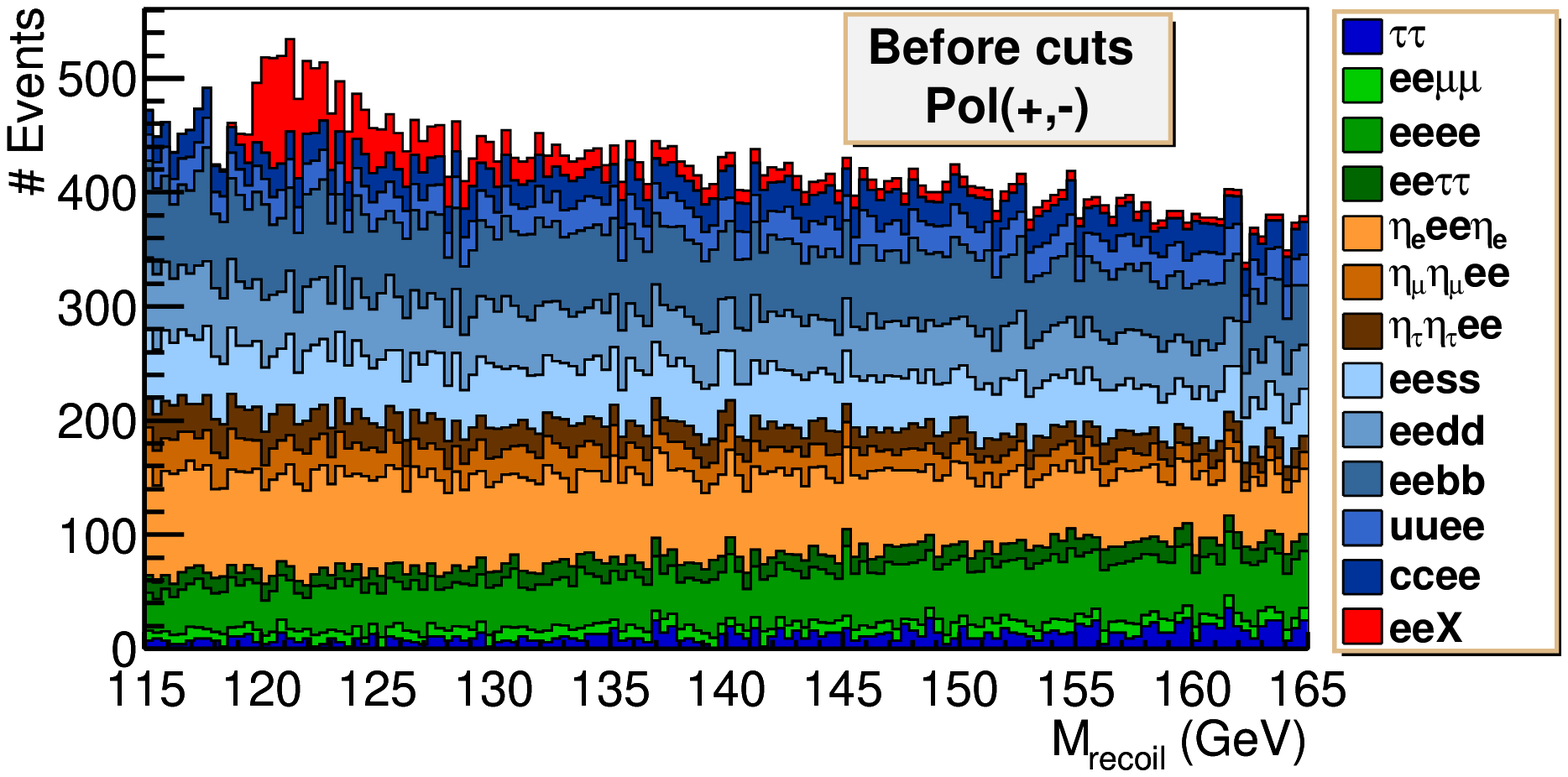} &
\includegraphics[width=6.7cm,height=4cm]{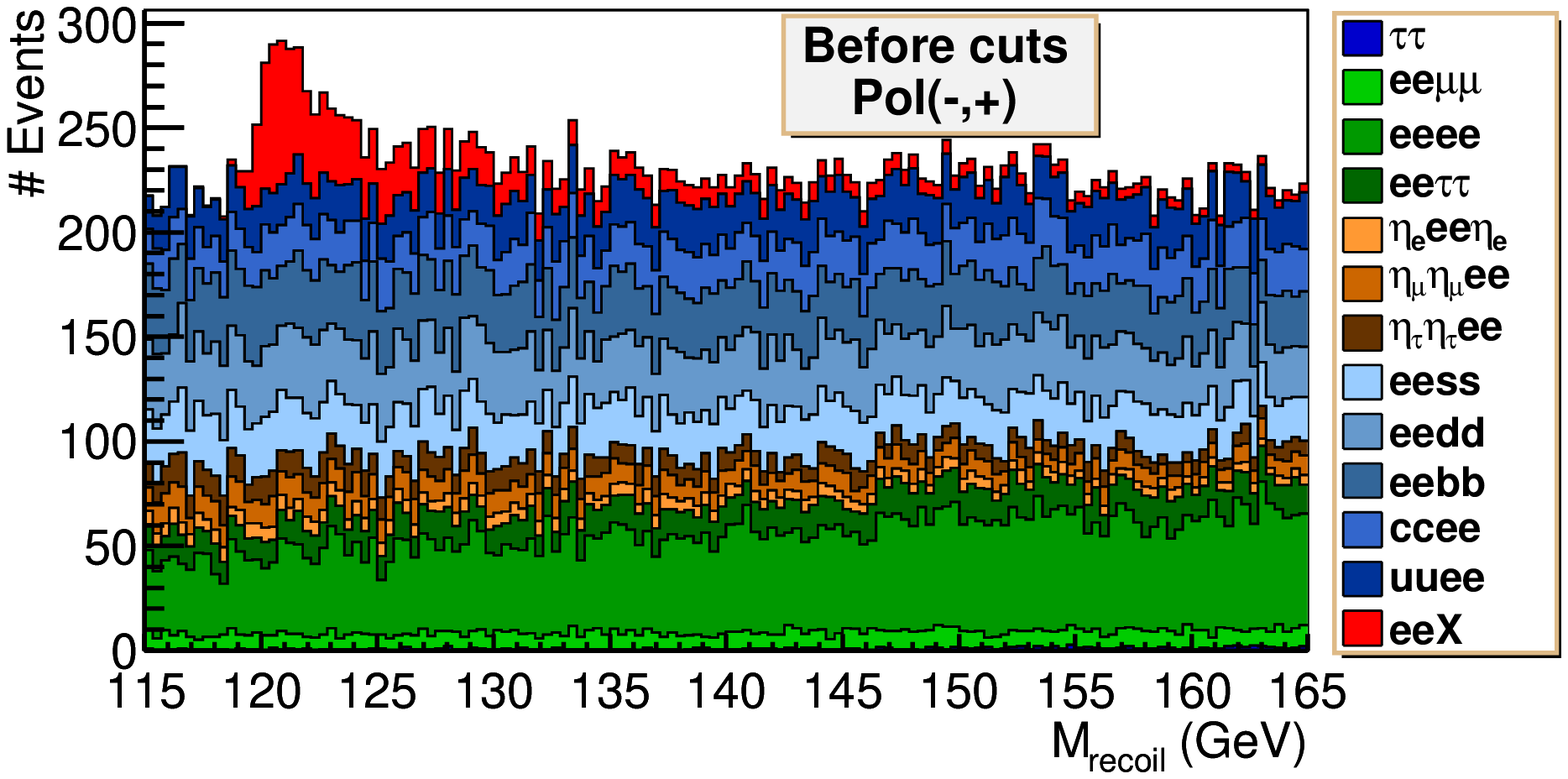}\\
\includegraphics[width=6.7cm,height=4cm]{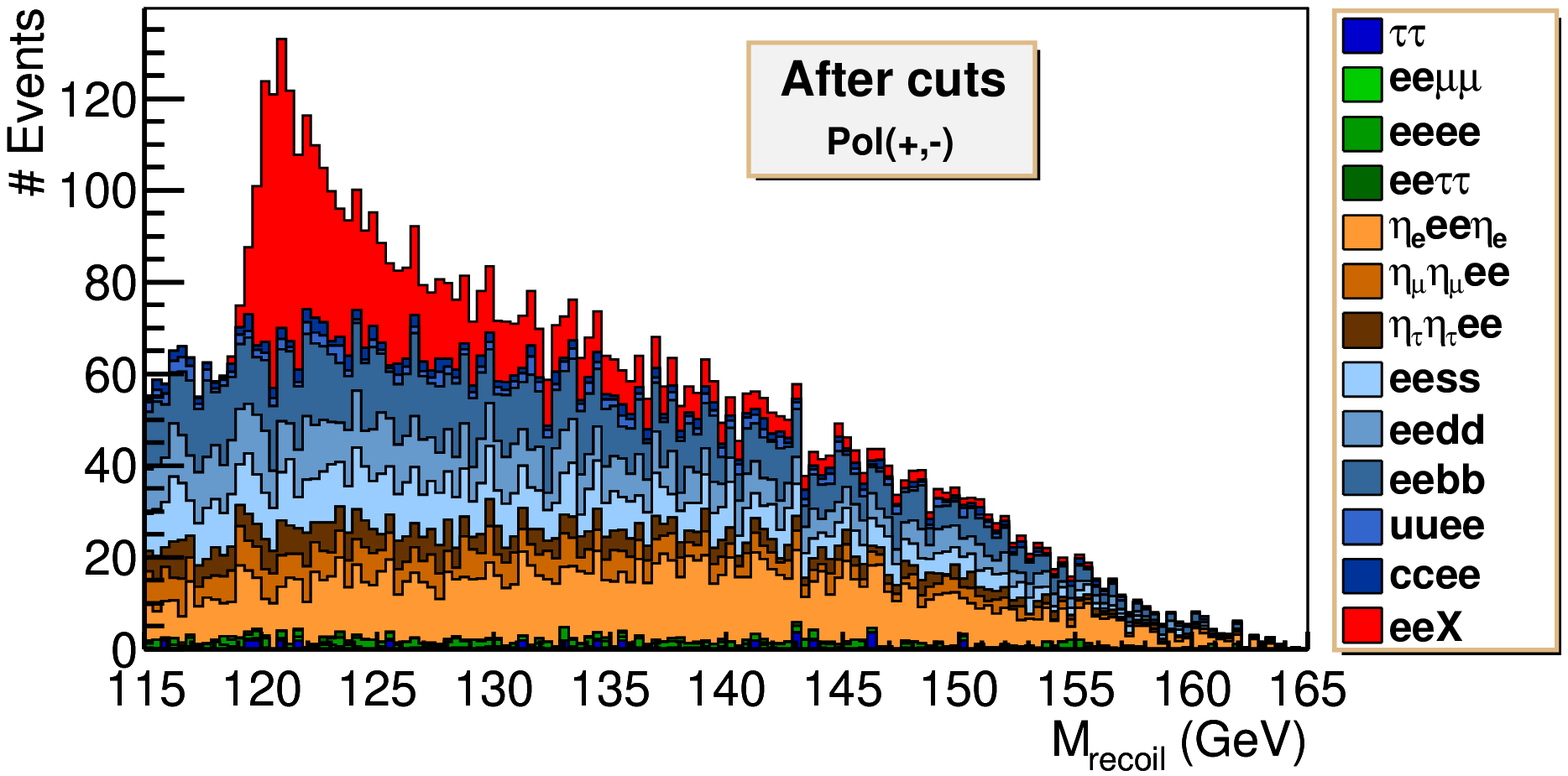}&
\includegraphics[width=6.7cm,height=4cm]{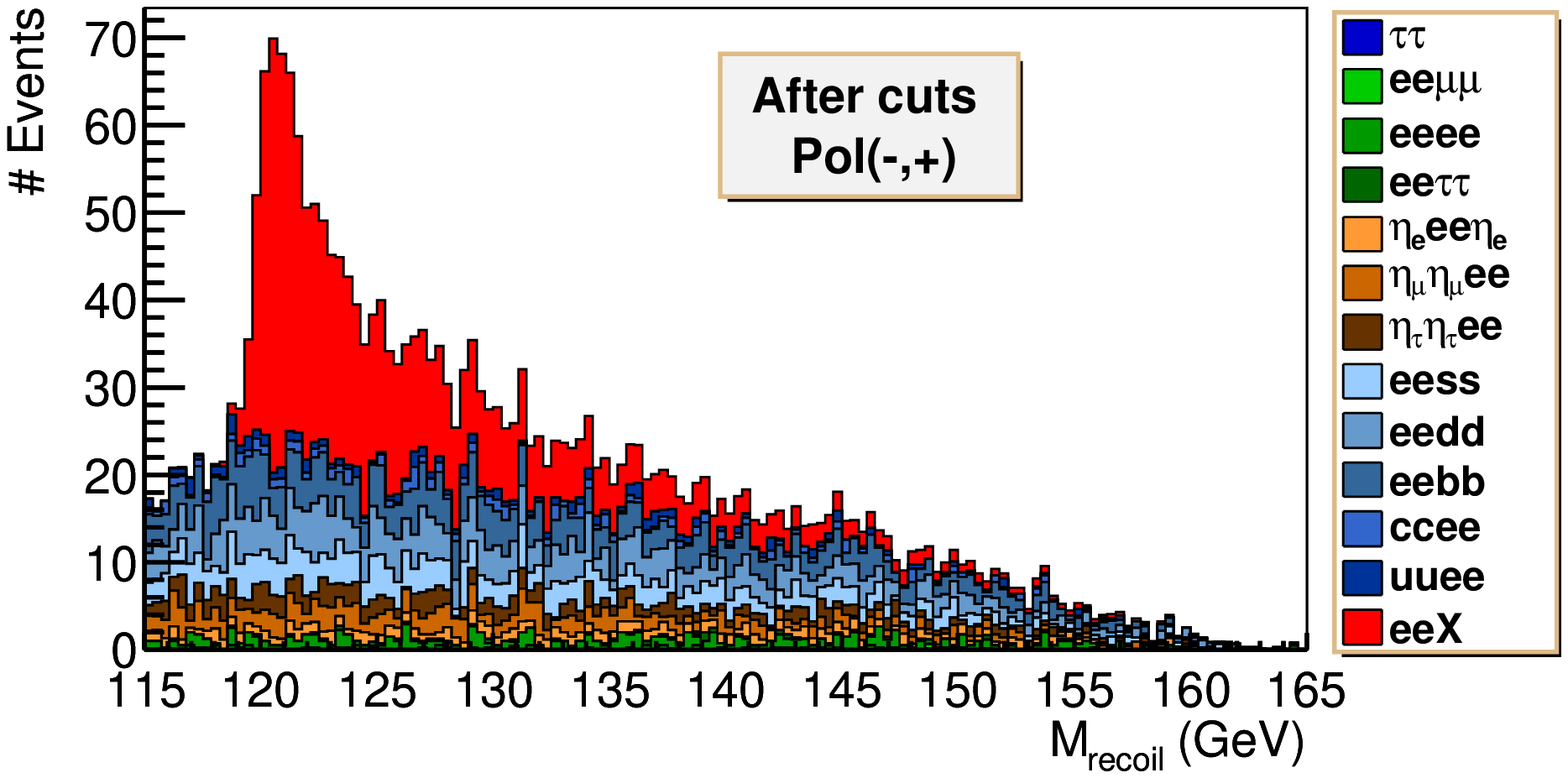}\\
\end{tabular}}
\caption{Reconstruted recoil mass distributions before and after the final cuts.}
\label{Fig:b_a_cuts}
\end{figure}

The following cuts were applied to reject the 4 fermion 
background: 
\begin{enumerate}
\item the di-lepton transverse momemtum calculated 
from the vectorial sum of the two leptons: 
18 GeV $<$ Pt$_{dilepton}$ $<$ 68 GeV,
\item the acollinearity (i.e. angle between the 2 leptons,
acol = cos$^{-1}$({\bf P}$_{1}$.{\bf P}$_{2}$/$\vert${\bf P}$_{1}$$\vert$.$\vert${\bf P}$_{2}$$\vert$) ): \\
0.4 $<$ acollinearity $<$ 1.35,
\item the energy of the Z boson: 
20 GeV $<$ E$_{Z}$ $<$ 115 GeV,
\item the Higgs mass interval: 
115 GeV $<$ M$_{H}$ $<$ 165 GeV.
\end{enumerate}
Different cuts are shown in Figure~\ref{Fig:cuts}.
Reconstruted recoil mass distributions before and after the selection 
are presented in Figure~\ref{Fig:b_a_cuts}.

\newpage

\onecolumn
\section{Reconstruction of the Higgs mass}
The Higgs recoil mass distribution is fitted with a GPET function 
("Gaussian Peak Exponential Tail") 
defined in the following way:
\begin{displaymath}
f(x; \alpha, n , \bar{x}, \sigma ) = N
 \left\{
 \begin{array}{ll}
 	e^{-\frac{\left(x - \bar{x}\right)^2}{2 \sigma^2}} & : for\ \frac{x-\bar{x}}{\sigma}\le\alpha 
\\
  	\beta \cdot e^{-\frac{\left(x - \bar{x}\right)^2}{2 \sigma^2}}
	+\left(1-\beta\right) \cdot 
	e^{\frac{\alpha^2}{2}} \cdot e^{-{\left(x - \bar{x}\right) \cdot \frac{\alpha}{\sigma}}} 
	& : for\ \frac{x-\bar{x}}{\sigma}>\alpha 
  \end{array}
 \right.
\end{displaymath}
where $\beta \in$ (0,1). This function and its first derivative are continuous 
when $(x - \bar{x})/\sigma = \alpha$.\\

When the cuts are applied to reject the background, one observes a small 
distorsion in the signal distribution in the region between 121 and 123 GeV. 
This can be removed by adding an additional Gaussian in the same region, 
as given by:
$$ f'(x; \alpha, n , \bar{x}, \sigma )~=~f(x; \alpha, n , \bar{x}, \sigma )~+~N_{1}G(x; \bar{x}_{1},\sigma_{1}) $$
The parameters $N_{1}$, $\bar{x}_{1}$ and $\sigma_{1}$ are determined in the fit ($\bar{x}_{1}$ is 
around 122 GeV and $\sigma_{1}$ is between 1 to 3 GeV). 
The corrected GPET function will be used in the following studies. \\   

\begin{figure}[h!]
  
   \begin{minipage}[b]{0.48\linewidth}
      \centering \includegraphics[width=6.9cm,height=5cm]{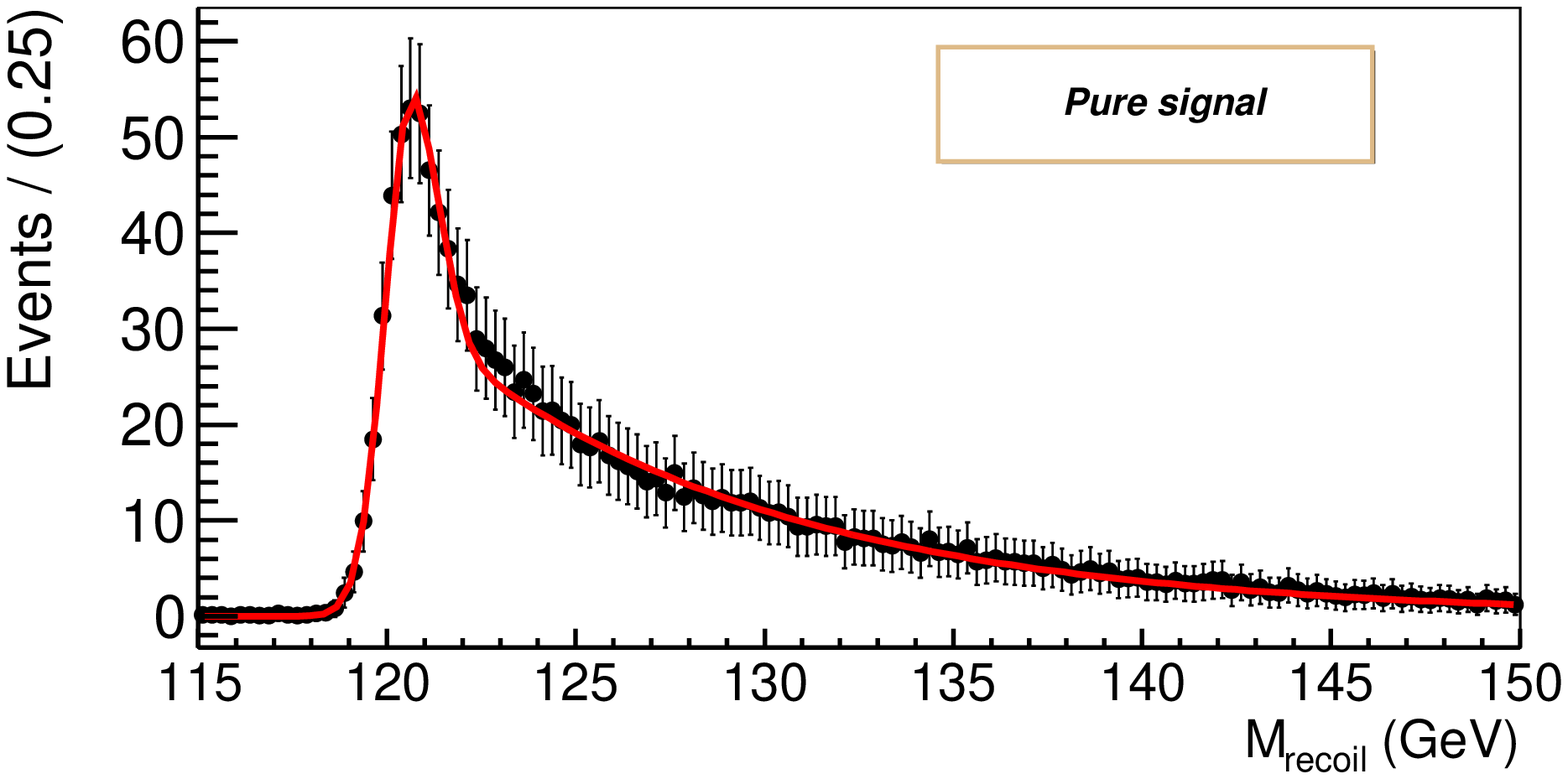}
      \centering \caption{Fit of the Higgs recoil mass distribution with the 
      corrected GPET function for the (+,-) beam polarization mode.}
      \label{Fig:pure_signal}
   \end{minipage}\hfill
   \begin{minipage}[b]{0.48\linewidth}   
      \centering \includegraphics[width=6.9cm,height=5cm]{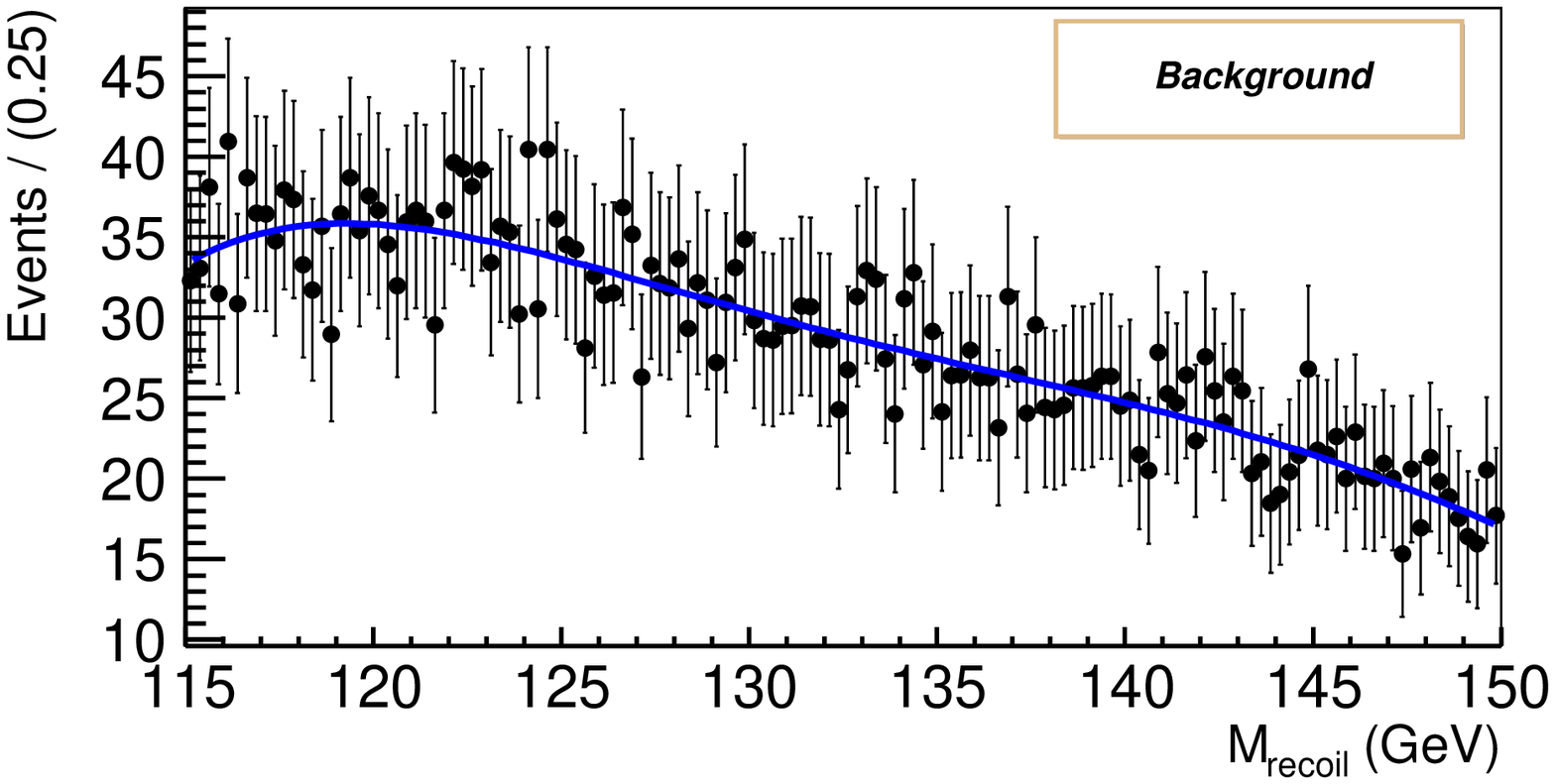}
      \centering \caption{Fit of the background distribution with a polynomial sum for the (+,-) beam polarization mode.}
      \label{Fig:background}
   \end{minipage}
\end{figure}

Figure~\ref{Fig:pure_signal} shows the pure signal 
fitted by the corrected function for the (+,-) polarization mode.\\

The background fits to a polynomial sum of degree 6, 
see Figure~\ref{Fig:background}.\\

\onecolumn
\section{Measurement of the Higgs mass with signal and background}
\begin{figure}[h]
%\vglue 2cm
\epsfxsize=6.9cm
\epsfysize=5cm
\begin{center}
\mbox{\epsffile{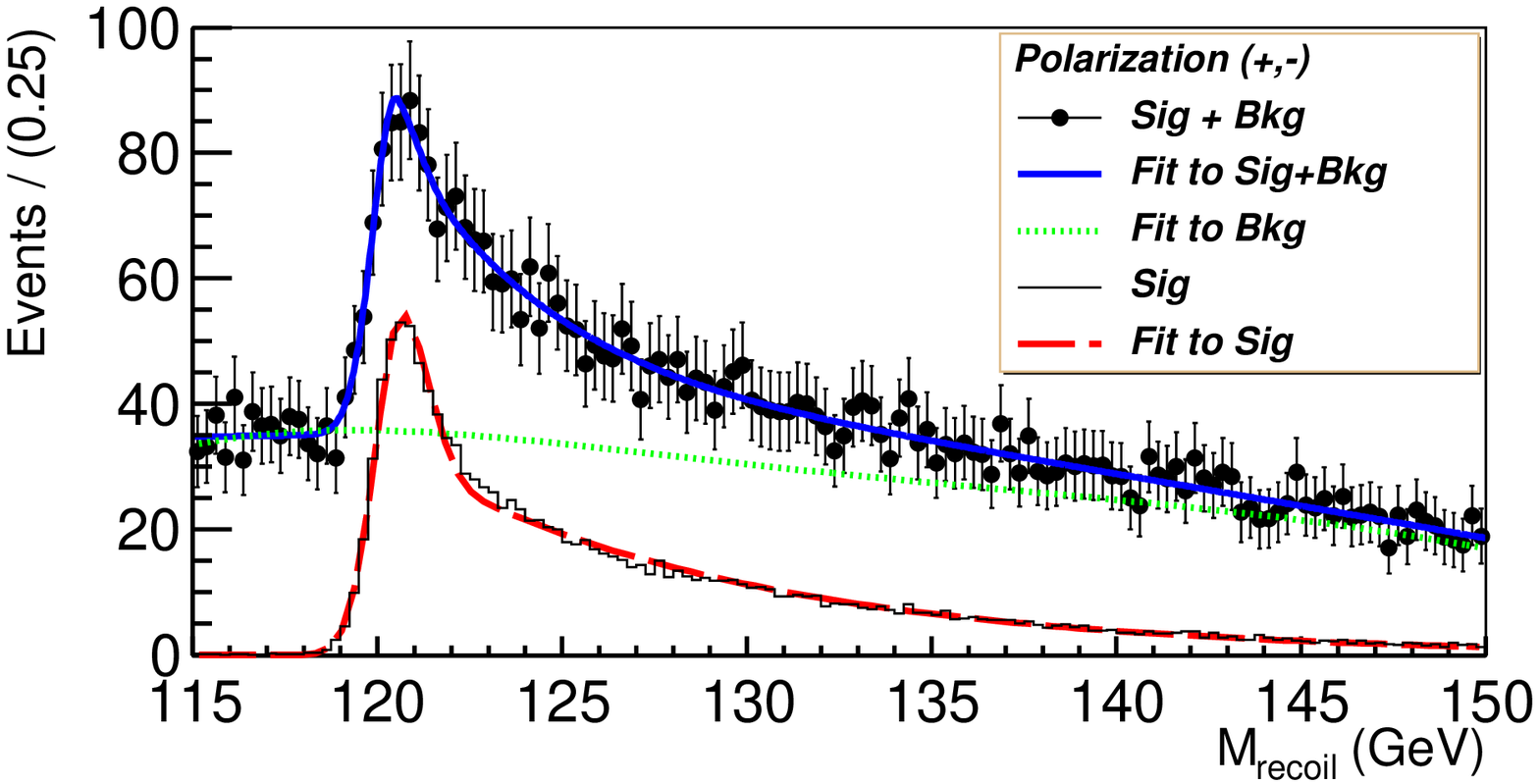}}
\mbox{\epsffile{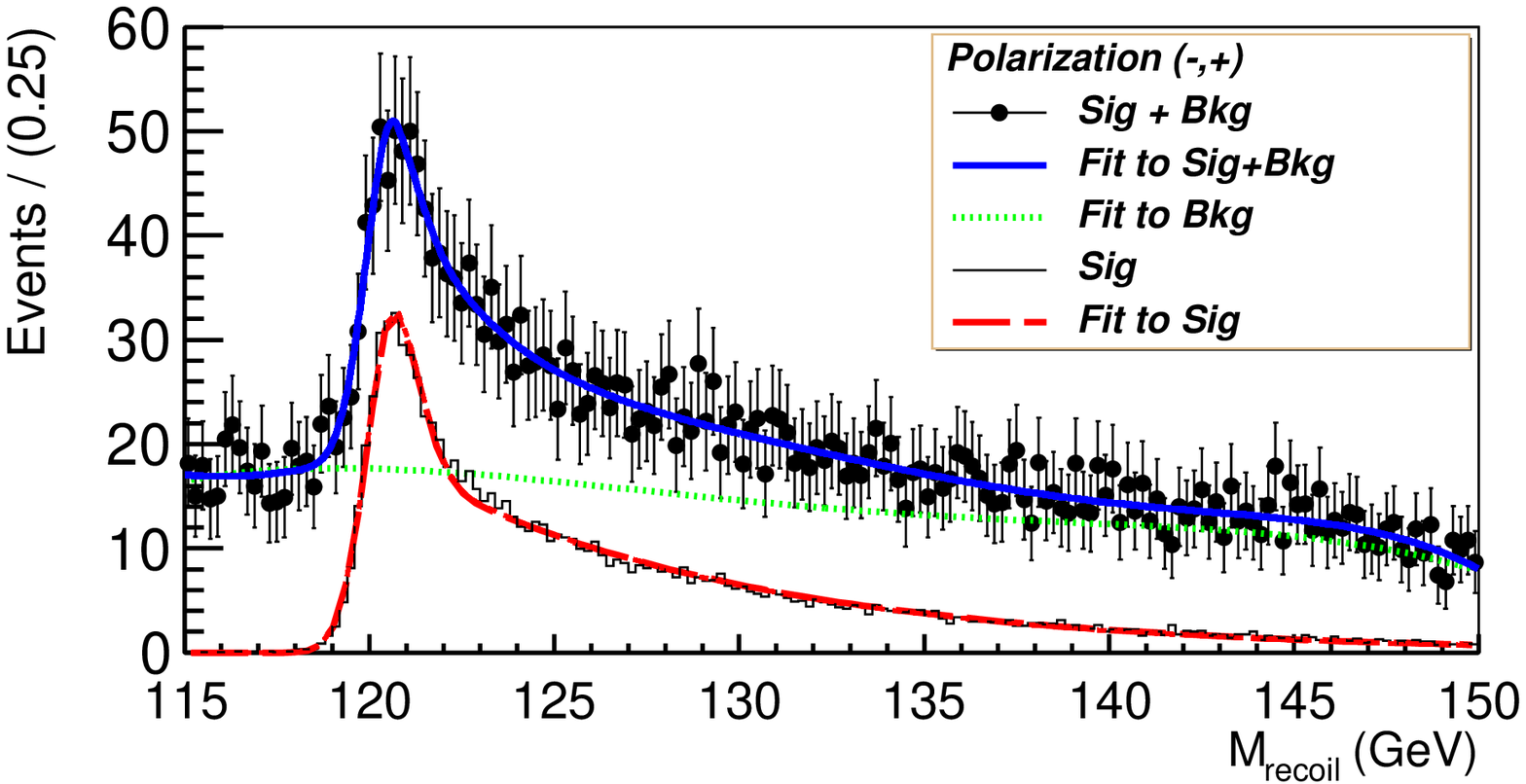}}
\caption [ ]
{Fits of the Higgs recoil mass distributions (signal + background)
for different beam polarization modes.}
\label{Fig:signal_bkgrd}
\end{center}
\end{figure}

Figure~\ref{Fig:signal_bkgrd} shows the Higgs recoil 
mass distributions (signal + background)
for different beam polarization modes.
Some of the fit parameters are given in Table~\ref{tab:Parameter_results}.
The accuracy on the Higgs 
mass measurement is around 100~MeV.

\vglue 0.5cm

\begin{wraptable}{l\onecolumn}{1.0\columnwidth}
\vglue 0.3cm
\renewcommand{\arraystretch}{1.4}
\centerline{\begin{tabular}{|c|c|c|c|c|}
   \cline{2-5}
   \multicolumn{1}{c|}{} & \multicolumn{4}{c|}{e$^{+}$e$^{-}$ beam polarization mode (30\%, 80\%)}\\
   \cline{2-5}
   \multicolumn{1}{c|}{} & \multicolumn{2}{c|}{(+,-)} & \multicolumn{2}{c|}{(-,+)}\\
   \cline{2-5}
   \multicolumn{1}{c|}{} & {Only signal} &{Signal + Bkgrd} & {Only signal} & {Signal + Bkgrd}\\
   \hline
   M$_{H}$ (GeV) & 120.486 $\pm$ 0.073 & 120.368 $\pm$ 0.100 & 120.507 $\pm$ 0.085 & 120.445 $\pm$ 0.110 \\
   \hline
   $\sigma$ (GeV) &  0.638 $\pm$ 0.051 &  0.575 $\pm$ 0.083 & 0.654 $\pm$ 0.062 & 0.592 $\pm$ 0.100 \\
   \hline
\end{tabular}}
\caption{Fit parameter results of the Higgs recoil mass 
for different polarization beam modes.}
\label{tab:Parameter_results}
\end{wraptable}

\onecolumn
\section{Conclusion and outlook}
The measurement of the Higgs mass recoil in the channel: 
e$^{+}$e$^{-} \to~$Z H $\to~$e$^{+}$e$^{-}$ + X 
illustrates the ILC potential for accurate measurements. 
The effect of the background is to deteriorate the accuracy on 
the Higgs boson mass. For M$_{H}$ = 120 GeV
and a luminosity of 250~fb$^{-1}$, it was found:\\
M$_{Rec}$ = 120.368 $\pm$ 0.100 GeV, for the beam polarization mode (+,-) and\\
M$_{Rec}$ = 120.445 $\pm$ 0.110 GeV, for the beam polarization mode (-,+).\\
The Higgs recoil mass distribution is asymmetric. 
This could be, for instance, partially corrected by taking into account 
the bremsstrahlung photons emmitted in the ILD detector and by 
using the electromagnetic calorimeter and
other sub-detectors.  
This is the next step in the analysis.\\
For more information please refer to~\cite{bibli:Li1,bibli:Li2}.
 
\section*{Acknowledgments}
The authors would like to thank Mr. H. Li for his help and his 
numerous advice.

% ****************************************************************************
% BIBLIOGRAPHY AREA
% ****************************************************************************

\begin{footnotesize}
% IF YOU DO NOT USE BIBTEX, USE THE FOLLOWING SAMPLE SCHEME FOR THE REFERENCES
% ----------------------------------------------------------------------------

% ----------------------------------------------------------------------------

\end{footnotesize}

% ****************************************************************************
% END OF BIBLIOGRAPHY AREA
% ****************************************************************************

\end{document}